\documentclass{article}
\usepackage{spconf,amsmath,graphicx}
\usepackage{booktabs}
\usepackage{multirow}
\usepackage[T1]{fontenc}
\usepackage{enumitem}
\setlist{nosep, leftmargin=14pt}

\usepackage{mwe} 

\title{DCE-FORMER: A TRANSFORMER-BASED MODEL WITH MUTUAL INFORMATION AND FREQUENCY-BASED LOSS FUNCTIONS FOR EARLY AND LATE RESPONSE PREDICTION IN PROSTATE DCE-MRI}
\name{
\begin{tabular}{@{}c@{}}
Sadhana S$^{1}$ \qquad Sriprabha Ramanarayanan$^{1,2}$ \qquad Arunima Sarkar$^{1}$ \\ Matcha Naga Gayathri$^{1}$ \qquad Keerthi Ram$^{2}$ \qquad  Mohanasankar Sivaprakasam$^{1,2}$
\end{tabular}
}
\address{$^{1}$ Indian Institute of Technology Madras (IITM), India \\ $^{2}$ Healthcare Technology Innovation Centre (HTIC), IITM, India}
\begin{document}
%
\maketitle
\begin{abstract}
Dynamic Contrast Enhanced Magnetic Resonance Imaging aids in the detection and assessment of tumor aggressiveness by using a Gadolinium-based contrast agent (GBCA). However, GBCA is known to have potential toxic effects. This risk can be avoided if we obtain DCE-MRI images without using GBCA. We propose, DCE-former, a transformer-based neural network to generate early and late response prostate DCE-MRI images from non-contrast multimodal inputs (T2 weighted, Apparent Diffusion Coefficient, and T1 pre-contrast MRI). Additionally, we introduce (i) a mutual information loss function to capture the complementary information about contrast uptake, and (ii) a frequency-based loss function in the pixel and Fourier space to learn local and global hyper-intensity patterns in DCE-MRI. Extensive experiments show that DCE-former outperforms other methods with improvement margins of +1.39 dB and +1.19 db in PSNR, +0.068 and +0.055 in SSIM, and -0.012 and -0.013 in Mean Absolute Error for early and late response DCE-MRI, respectively.
\end{abstract}
\section{Introduction}
\label{sec:intro}
The Dynamic Contrast Enhanced Magnetic Resonance Imaging (DCE-MRI) sequence produces a visualization of angiogenesis and highlights subtle tumor lesions. The increased permeability of tumor vessels is imaged with a contrast agent such as Gadolinium (Gad) which washes in and out of the tumor more rapidly than normal tissue. 
 DCE-MRI involves early-phase and late-phase contrast enhanced imaging to capture the morphological characteristics of the tumor over time.  
However, there are safety concerns around Gad retention and its potential toxic effects \cite{moshe_gbca_toxicity}. It is beneficial to reduce the dosage or avoid the Gad contrast agent.
\noindent The existing DCE-MRI involves anatomic acquisitions such as T1 pre-contrast and T2-Weighted (T2W) MRI, along with Apparent Diffusion Coefficient (ADC), which quantifies water molecule diffusion within the tissue.
The purpose of this work is to synthesize early and late-phase DCE-MRI given such non-contrast images which capture structural and perfusion information.

Synthetic MRI, enabled by deep learning, has been explored for reducing or eliminating Gad-based imaging in brain cancer \cite{corey_HR_convNet}. 
Deep Generative Adversarial Networks (GAN) is a prominent method with prior works exploring low-dose and cross-modality MRI synthesis \cite{fujita_DL_medimg_analysis}.
In \cite{gong_dl_red_gddose}, a deep learning model is proposed for generating full-dose late response images from both pre-contrast and low-dose images for brain MRI. A conditional GAN with a tailor-made contrast enhancement loss function \cite{fonnegra_earlytolate} is formulated to generate late response from early response in Breast MRI images. 
TSGAN \cite{cho_tsgan} focuses on synthesizing DCE-MRI images from non-contrast MRI images.
\begin{figure}
    \centering
    \includegraphics[width=1\linewidth]{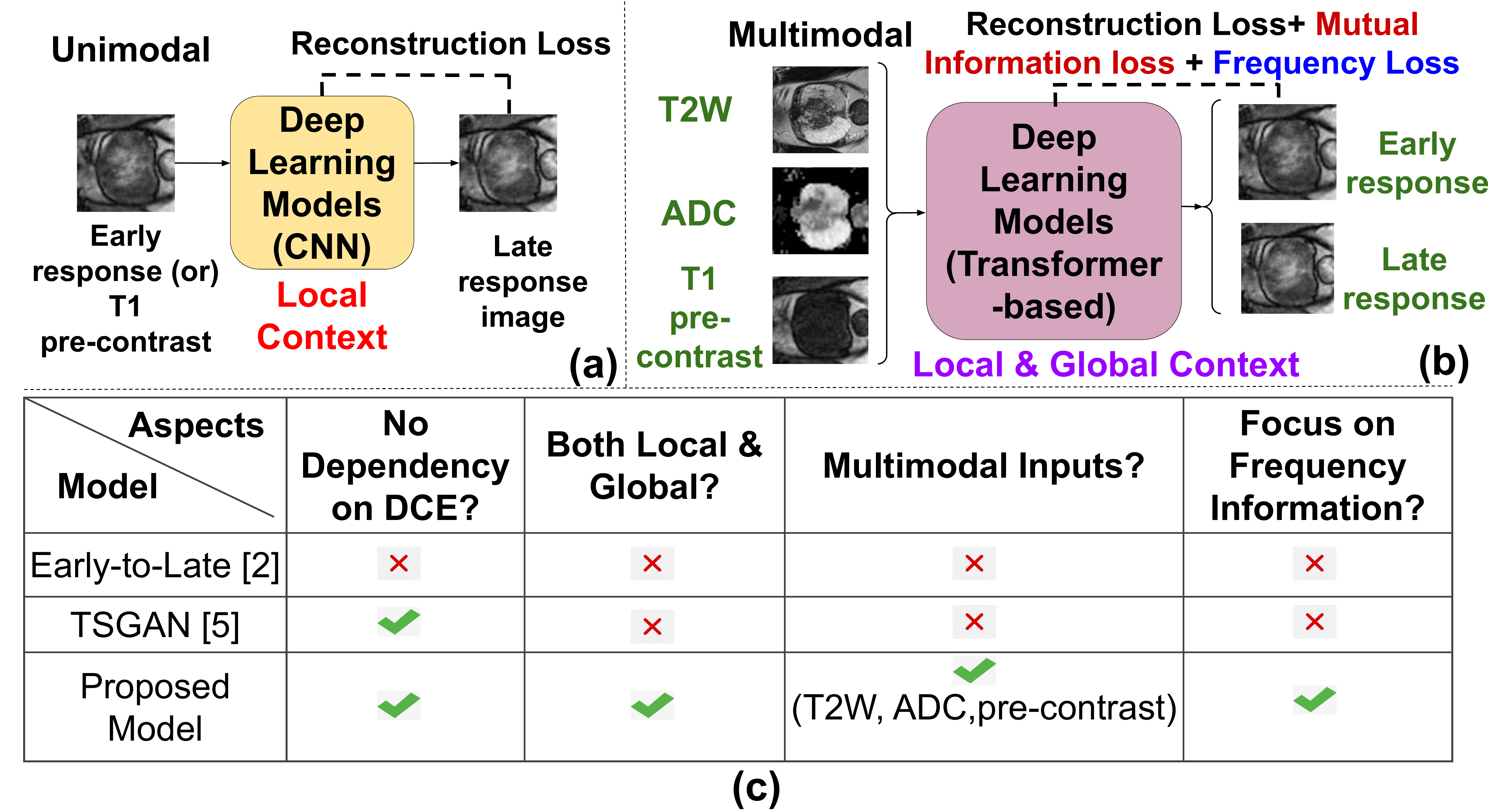}
    \caption{Concept diagram of DCE-MRI synthesis approaches \textbf{(a)} Previous methods focus on local contextual learning from early response using CNNs and without ADC information. \textbf{(b)} \& \textbf{(c)} The proposed DCE-Former is a transformer-based network that learns both local and global pixel dependencies, captures semantic details and contrast-enhancing patterns using mutual information and frequency-based loss functions.}
    \label{fig:concept diagram}
\end{figure}
Inspired by the multimodal learning capabilities to capture salient, mutual, and complementary information from multiple modalities, the encoder-decoder convolutional neural network (CNN) \cite{tsaftaris_mm_mrsynth} \cite{corey_HR_convNet} and Hi-Net \cite{zhou_hinet} synthesize brain MR images by fusing the latent representations of multimodal MR images.  
\begin{figure*}
    \centering
    \includegraphics[width=1\linewidth]{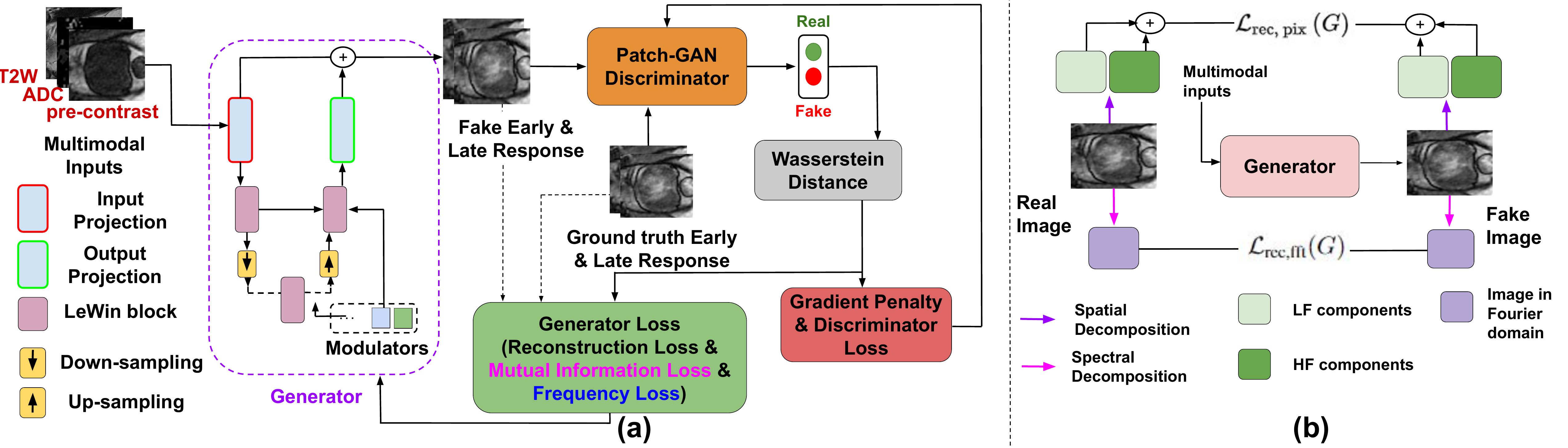}
    \caption{ \textbf{(a)} DCE-former architecture consists of Uformer as the generator and PatchGAN discriminator with the two loss components (MI-based and frequency-based) to optimize the model weights. \textbf{(b)} The proposed frequency loss function computes the loss in the low and high-frequency components in the spatial and spectral domains.}
    \label{fig:architecture diagram}
\end{figure*}

Unlike prior works (Fig.\ref{fig:concept diagram}), we consider the problem of synthesizing both early and late response images of DCE-MRI from multimodal non-contrast inputs by using a transformer-based model to capture local and global contrast details as the dynamics of the contrast in DCE acquisitions might not be localized to a particular region within the prostate.
In particular, we use Uformer \cite{wang_uformer} which offers core benefits making it extensible to image synthesis tasks, namely (i) the Locally Enhanced Window (LeWin) transformer blocks, and (ii) multi-scale restoration modulator to capture local and global dependencies to generate details from all feature locations at hierarchical levels synergistically.
Furthermore, to effectively combine the physiological features that involve time-varying contrast-enhancing patterns of normal and diseased tissue types, we use two loss components (i) mutual information-based loss to capture nonlinear statistical dependencies between multimodal inputs and predicted images, distinct from the approach in \cite{bengio_mine}, and (ii) frequency-based loss function in both the spatial and Fourier domains \cite{fdit} to preserve the contrast levels of the DCE-MRI images and finer details like local hyper-enhancement of intensities within lesions.We summarize our contributions as,
 \begin{enumerate}
     \item We propose DCE-former, a transformer-based architecture, in a GAN network to synthesize early and late-phase prostate DCE-MRI images from multimodal inputs such as T2W, ADC, and T1 pre-contrast images.
     \item We propose two loss functions, (i) a mutual information-based loss function to capture the complementary information about contrast uptake in DCE-MRI, and (ii) a frequency-based loss function in the pixel and Fourier space to learn local and global hyper-intensity patterns.
     \item Extensive experiments, using ProstateX dataset, show that the proposed DCE-former predicts the early and late response DCE-MR images with improvement margins of  (i) +1.39 dB PSNR, +0.068 SSIM, -0.012 MAE, -18.00 FID for early response, and (ii) +1.19 db PSNR, + 0.055 SSIM, -0.013 MAE, -15.29 FID for late response, over the second best-performing baseline model, ResViT.
 \end{enumerate}
\begin{table*}[t]
\caption{Quantitative Comparison of the synthesized early \& late DCE-MRI timepoints between DCE-former \& other models}
\label{tab:results table}
\resizebox{\textwidth}{!}{%
\begin{tabular}{@{}ccccccccc@{}}
\toprule
\multirow{2}{*}{\textbf{Model}} & \multicolumn{4}{c}{\textbf{Early Response}}                                                         & \multicolumn{4}{c}{\textbf{Late Response}}                                                         \\ \cmidrule(lr){2-5} \cmidrule(lr){6-9} 
                                & \textbf{PSNR}$\uparrow$           & \textbf{SSIM}$\uparrow$              & \textbf{MAE}$\downarrow$    & \textbf{FID Score}$\downarrow$ & \textbf{PSNR}$\uparrow$            & \textbf{SSIM}$\uparrow$             & \textbf{MAE}$\downarrow$    & \textbf{FID Score}$\downarrow$ \\ \midrule
ConvLSTM                        & 14.92 +/- 1.50         & 0.2397 +/- 0.04         & 0.1393          & 118.706           & 15.27 +/- 2.56           & 0.2392 +/- 0.06         & 0.1351          & 115.480             \\
Pix2Pix                         & 15.21 +/- 5.49         & 0.2886 +/- 0.18          & 0.1121          & 65.8695            & 13.80 +/- 5.05           & 0.2321 +/- 0.19         & 0.1288          & 32.8592            \\
ResViT                          & 21.46 +/- 0.04        & 0.6369 +/- 0.04           & 0.0638          & 32.4624            & 20.88 +/- 0.04         & 0.6231 +/- 0.05         & 0.0684          & 30.0611            \\
RegGAN                          & 20.56 +/- 0.02       & 0.5966 +/- 0.02          & 0.0571          & 23.7963            & 20.09 +/- 0.02        & 0.5803 +/- 0.02         & 0.0617          & 22.6124            \\
TSGAN                            & 21.16 +/- 3.50         & 0.6253 +/- 0.10           & 0.0693          & 23.7533            & 20.46 +/- 2.64          & 0.5926 +/- 0.09         & 0.0749          & 24.6665            \\
\textbf{DCE-former}             & \textbf{22.85 +/- 3.74} & \textbf{0.7047 +/- 0.11} & \textbf{0.0522} & \textbf{14.4614}   & \textbf{22.07 +/- 3.33} & \textbf{0.6790 +/- 0.13} & \textbf{0.0558} & \textbf{14.7617}   \\ \bottomrule
\end{tabular}%
}
\end{table*}

\vspace{-1.5pt}
\section{Methodology}
\label{sec:methodology}
\subsection{Early and Late response using GANs}
\label{ssec:Early and Late response using GANs}
The image synthesizing task can be formulated using GANs by coercing the generated post-contrast early and late response images to be as plausible as the ground truth post-contrast DCE images while learning the spatial coherence, tissue dynamics, and perfusion information from the input  $\tilde{X}$. $\tilde{X}$ is concatenation of T2W, ADC and T1 pre-contrast images ($\tilde{X}=X_{T 2, A D C, \text { pre. }}$). This adversarial approach is given by,
\begin{equation}
G(\tilde{X})^*=\arg \min _G \max _D\left(G(\tilde{X}), D\left(Y_{e, l}\right)\right) L_{G A N} \text {. }
\end{equation}
\subsection{Mutual Information (MI) Loss}
\label{ssec:Mutual Information Loss}
The reconstruction loss component ($\mathcal{L}_{1}$), which computes pixel-wise differences, might not capture the perceptually relevant characteristics of the image, like textural details in the time-varying contrast patterns \cite{ledig_photorealistic_imgSR}.
To drive the model to generate temporal contrast enhancement patterns coherently, we use the MI loss function to learn representations that yield high semantic similarity 
with regard to the underlying tissue information between the real and generated images. The MI loss, $\mathcal{L}_{\text {MI }}$, between the real early (and late) response images and the generated early (and late) response is given by,
\begin{multline*}
MI\left(X_{i} ; G(\tilde{X})\right)=H\left(X_{i}\right)+H\left(G(\tilde{X})\right)-H\left(X_{i}, G(\tilde{X})\right)
\end{multline*}
\begin{equation}\label{4}
\mathcal{L}_{\text {MI }}(G)=\frac{2 * MI\left(X_{i} ; G(\tilde{X})\right)}{H\left(X_{i}\right)+H\left(G(\tilde{X})\right)}
\end{equation} 
Here, $X_{i}$ and $G(\tilde{X})$ are the real and fake DCE-MR images, where $i$ denotes the early (or late) response image. $H\left(X_{i}\right)$ and $H\left(G(\tilde{X})\right)$ denote the individual entropy of the images and $H\left(X_{i}, G(\tilde{X})\right)$ denotes the joint entropy. Our main objective is to maximize the mutual information between the ground truth and generated early and late response images.
\subsection{Frequency Loss}
\label{ssec:Frequency Loss}
The frequency loss function \cite{fdit} aids in learning local and global features at the optimization level as (i) Convolution by Gaussian kernel in spatial domain represents frequency features in a local manner (ii) Discrete Fourier Transform (DFT) characterizes frequency distribution globally as each spatial frequency is computed by employing information from all pixels. 
The high-frequency (HF) component, $X_H$, of the image, which preserves fine-grained contrast details, is computed from the low (LF) component, $X_L$, which is obtained by applying a Gaussian kernel, as follows.
\begin{equation}
X_H=X-X_L
\end{equation}
The spatial domain reconstruction loss term for both low-frequency and high-frequency components is given by,
\begin{equation}
\begin{aligned}
\mathcal{L}_{\text {rec,pix }}(G)=E[\| & X_L-(G(X))_L \|_1 \\
& \left.+\left\|X_H-(G(X))_H\right\|_1\right]
\end{aligned}
\end{equation}
For the frequency domain loss function, we apply the DFT ($F$) on the image. The frequency spectrum reconstruction loss is regulated as,
\begin{equation}
\mathcal{L}_{\text {rec,fft}}(G)=E\left[\left\|F^R(X)-F^R(G(X))\right\|_1\right]
\end{equation}
The overall loss is formulated as,
\begin{multline*}
G^*=\arg \min _G \max _D(G, D) L_{G A N}+ \lambda_1\mathcal{L}_{1}+ \\ \left(1-\lambda_{\text {MI }} \mathcal{L}_{\text {MI }}(G)\right) +\lambda_{\text {rec,pix }} \mathcal{L}_{\text {rec, pix }}(G)+\lambda_{\text {rec, fft }} \mathcal{L}_{\text {rec,fft }}(G)
\end{multline*}

where $\lambda_1$ , $\lambda_{\text{MI}}$ , $\lambda_{\text {rec,pix }}$ and $\lambda_{\text {rec, fft }}$ are the weight hyper-parameters for each component of the loss function. 
\section{DATASET \& IMPLEMENTATION DETAILS}
\label{sec:DATASET IMPLEMENTATION DETAILS}
\begin{figure*}
    \centering
    \includegraphics[width=1\linewidth]{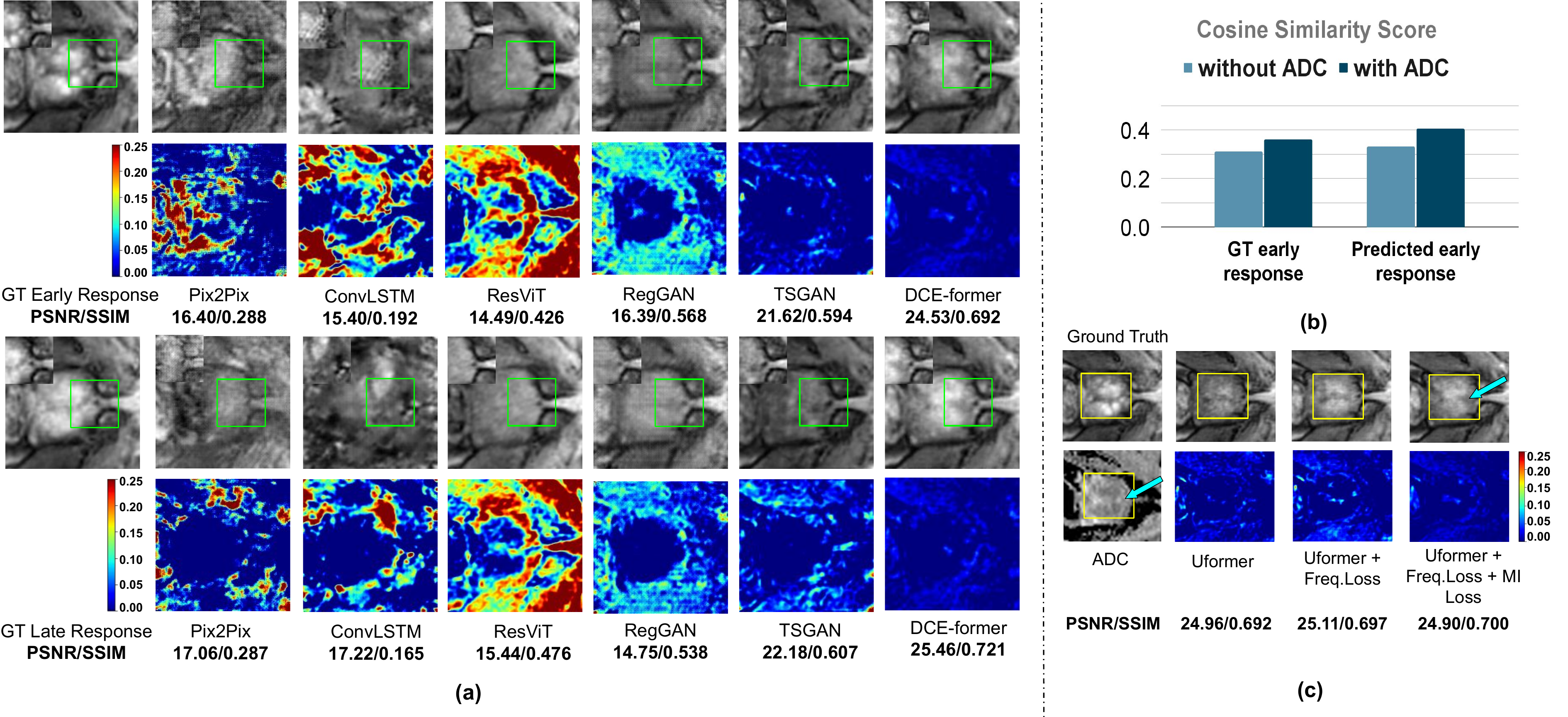}
    \caption{ (a) Visual results of early \& late response (from left to right): Ground truth, Pix2Pix, ConvLSTM, ResViT, RegGAN, TSGAN, and  DCE-former (ours) (b) Comparison of similarity scores between predicted \& ground truth early response images with \& without ADC in multimodal input.(c) Qualitative results for the ablative study. A combination of both MI and frequency loss effectively captures the contrast enhancement patterns. The DCE-MRI predictions image shows matching and complementary features with respect to the ADC (as shown by blue arrows). Green and yellow boxes denote the regions of interest.}
    \label{fig:qualitative results}
\end{figure*}
We have used ProstateX dataset \cite{prostatex} which consists of the 346 patients' studies acquired without an endo-rectal coil consisting of 5520 images, (4416 and 1104 images for training and validation respectively). Each patient data consists of T2W, ADC, T1 pre-contrast and DCE-MRI images. We used SimpleITK rigid registration framework to align all the images. The prostrate organ is cropped to a size of 160 $\times$ 160 $\times$ 16.

The proposed DCE-former (in Fig.\ref{fig:architecture diagram}) is a Wasserstein GAN with Gradient Penalty (WGAN-GP) with Uformer \cite{wang_uformer} as the generator and a Patch-GAN discriminator. We have compared DCE-former with five models, namely Pix2Pix \cite{isola_i2i_cgan}, ConvLSTM \cite{shi_convlstm}, ResViT \cite{dalmaz_resvit}, RegGAN \cite{kong_reggan} and TSGAN \cite{cho_tsgan}.
We fixed the middle time point as the early response image and the last time point as the late response image in the ground truth DCE-MRI sequences with varying numbers of time points across studies. In the transformer network, we have fixed the number of LeWin blocks as \{1, 2, 8, 8\} in the corresponding levels of the encoder and the decoder. The optimal Gaussian kernel size (for the frequency loss) is 13$\times$ 13 for our curated dataset. 
All models are implemented in Pytorch 2.0 and are trained for 200 epochs, batch size = 6, and hyperparameters, $\lambda_1$ = 5, $\lambda_{M I}$ = 10, $\lambda_{\text {rec,pix }}$ = 10 and $\lambda_{\text {rec, fft }}$ = 10.
The evaluation metrics are PSNR, SSIM, MAE (Mean Absolute Error), and FID (Frechet Inception Distance).
\section{results}
\label{sec:results}
Our experiments include (i) a comparative study against other GAN methods for contrast translation, (ii) an ablative study of the various loss components used in our approach and (iii) an analysis of the contribution of ADC images in synthesizing DCE-MR images. Table \ref{tab:results table} shows the quantitative comparison of the proposed approach with other methods. Our observations are:  (i) Our model (for both early and late response images) outperforms all the other methods consistently across all the evaluation metrics. (ii) Our model also performs better than the standard MRI image-to-image translation benchmark, ResViT. Visual comparison in Fig.\ref{fig:qualitative results}(a) shows that our DCE-former is able to generate the hyperintensity patterns in the early and late response images very close to the ground truth as compared to other methods. We believe that the reason for this observation is that our approach learns essential details from ADC images which have the potential to reflect the perfusion properties of tissues. Furthermore, the structural correlation between the multimodal inputs and the outputs is learned using T2W and T1 pre-contrast MRI images. (iii) Although TSGAN generates images with comparatively less residual error, it is unable to capture the exact contrast behavior due to its limited capability of long-range dependencies. 
Our DCE-former overcomes this limitation by leveraging global and local contextual learning at both architecture and optimization levels.
(iv) MI helps to capture the matching and complementary features from the multimodal inputs during optimization in synthesizing DCE-MRI images. Aided by frequency loss, the fine-grained contrast details are learned in both spatial and frequency domains.
From Figure \ref{fig:qualitative results}(c), we see that the matching complementary features refer to dark and bright (hyperintense) regions in the ADC and DCE-MRI images, respectively. 
Wilcoxon signed-rank test shows that our metrics are statistically significant (p < 0.05).

\vspace{-0.5cm}
\subsection{Ablative Studies}
\noindent\textbf{Ablative Study with ADC:} Figure 3(b) emphasizes the importance of learning from ADC images during training. We note that the similarity for real (and synthesized) DCE-MRI images is higher when ADC is used as input compared to the case with inputs without ADC. This is due to the strong correlation between DCE-MRI and ADC images.

\noindent\textbf{Ablative Study of the loss components:} In this study, we understand the contributions of optimizing the network using each of the three loss functions. From Table 2, we note that the L1 loss which minimizes pixel-wise differences is inadequate in capturing physiological details. Both MI and DFT loss components play a key role in improving the overall perceptual quality of the prediction and a suitable combination of both components improves the quality further. The visual results in Figure \ref{fig:qualitative results}(c) affirm these observations qualitatively.  
\begin{table}[t]
\caption{Ablative study on the loss components}
\label{tab:ablative table}
\resizebox{\columnwidth}{!}{%
\begin{tabular}{@{}ccc@{}}
\toprule
\multirow{2}{*}{\textbf{Model}}                                                                   & \multicolumn{2}{c}{\textbf{DCE-MRI output}} \\ \cmidrule(l){2-3} 
                                                                                                  & \textbf{PSNR}       & \textbf{SSIM}         \\ \midrule
$\mathcal{L}_{1}$                                                                                 & 23.01 +/- 3.72     & 0.6880 +/- 0.113      \\
$\mathcal{L}_{1}$ + $\mathcal{L}_{\text {rec,pix}}$ + $\mathcal{L}_{\text {rec,fft}}$             & 23.08 +/- 3.73     & 0.6919 +/- 0.113     \\
$\mathcal{L}_{1}$+$\mathcal{L}_{\text {rec,pix}}$ + $\mathcal{L}_{\text {rec,fft}}$ + $\mathcal{L}_{\text {MI}}$ & 23.03 +/- 3.44     & 0.7047 +/- 0.119     \\ \bottomrule
\end{tabular}%
}
\end{table}

\vspace{-1.5pt}
\section{CONCLUSION }
\label{sec:CONCLUSIONS}
We propose DCE-former, transformer-based model in a GAN setting to synthesize early and late response images in DCE-MRI from multimodal inputs such as T2W, ADC, and T1 pre-contrast images. To generate post-contrast patterns, we establish an MI loss and frequency-based loss which leverage the global and local contextual learning of the transformer to capture the physiological features and complementary information about contrast intake. We are currently extending our approach to predict pharmaco-kinetic parametric maps to estimate various physiological parameters in DCE-MRI.
\section{Compliance with ethical standards}
\label{sec:ethics}
This research study was conducted retrospectively using human subject data made available in open access by the Cancer Imaging Archive. Ethical approval was not required as confirmed by the license attached with the open-access data. 
\section{Acknowledgments}
\label{sec:acknowledgments}
Acknowledgement withheld.

\bibliographystyle{IEEEbib}
\bibliography{strings,refs}

\end{document}